\documentclass[twocolumn,aps,prl,amssymb,superscriptaddress]{revtex4-2}

\usepackage{graphicx}
\usepackage{lipsum}
\usepackage{float}
\usepackage{amsmath}
\usepackage{mathtools}
\usepackage{bbold}
\usepackage{physics}
\usepackage{ulem}
\usepackage{siunitx}
\usepackage[dvipsnames]{xcolor}

\usepackage[utf8]{inputenc}
\usepackage{comment}

\usepackage[breaklinks=true]{hyperref}
\hypersetup{
        colorlinks = true,
        urlcolor = blue,
        citecolor = blue
}

\begin{document}

\title{Aharonov-Bohm interferometer in inverted-band pn junctions} 

\author{Yuhao Zhao}
\email{yzhao.phy@gmail.com}
\affiliation{\mbox{Institute for Theoretical Physics, ETH Zurich, 8093 Zurich, Switzerland}}
\author{Oded Zilberberg}
\affiliation{\mbox{Department of Physics, University of Konstanz, D-78457 Konstanz, Germany}}
\author{Antonio \v{S}trkalj}
\email{astrkalj@phy.hr}
\affiliation{\mbox{Department of Physics, Faculty of Science, University of Zagreb, Bijeni\v{c}ka c. 32, 10000 Zagreb, Croatia}}

\begin{abstract}
    Inverted-band $pn$ junctions in two-dimensional materials offer a promising platform for electron optics in condensed matter, as they allow to manipulate and guide electron beams without the need for spatial confinement. In this work, we propose the realization of an Aharonov-Bohm (AB) interferometer using such $pn$ junctions. 
    We observe AB oscillations in numerically obtained conductance and analytically identify the conditions for their appearance by analyzing the scattering processes at the $pn$ interface.
    To support experimental implementation, we also consider junctions with a graded interface, where the potential varies smoothly between the $p$- and $n$-doped regions. Our results reveal an abrupt change in the AB-oscillation frequency, which we attribute to a distinct transition in the hybridization across the interface. We verify that our reported AB oscillations are robust to realistic disorder and temperature decoherence channels.
    Our study paves the way for realizing the Aharonov-Bohm effect in bulk mesoscopic systems without the need for external spatial confinement, offering new possibilities in electron optics.
\end{abstract}
\maketitle

The Aharonov-Bohm (AB) effect is one of the cornerstone phenomena highlighting the non-local nature of quantum mechanics~\cite{ehrenberg_refractive_1949,aharonov_significance_1959,aharonov_further_1961}. The effect arises when two quantum trajectories enclose a finite magnetic flux, making their interference dependent on the external magnetic field~\cite{ihn_semiconductor_2009}. On a microscopic scale, the AB effect has been observed in a wide range of systems, and plays a pivotal role in revealing quantum phenomena in transport that strongly depends on magnetic fields, e.g. in revealing weak localization through magnetoresistance measurements~\cite{Akkermans_Montambaux_2007,chen_weak_2019}.

To directly realize an AB interferometer in a mesoscopic system, a variety of two-dimensional platforms have been employed~\cite{van_oudenaarden_magneto-electric_1998,bachtold_aharonovbohm_1999,russo_observation_2008,kremer_square-root_2020,iwakiri_tunable_2024,kawaguchi_pseudo-spin_2024, kim_aharonovbohm_2024,biswas_anomalous_2024}, where clear oscillations in transport as a function of an external magnetic field were observed. The primary challenges in observing such AB oscillations lie in the ability to prepare well-defined electron paths while maintaining coherence between electrons traveling along them~\cite{van_oudenaarden_magneto-electric_1998, zilberberg2011charge}. 
Guiding electrons in such a manner is inherently difficult and has traditionally been achieved through spatial confinement using etching techniques~\cite{russo_observation_2008,kim_aharonovbohm_2024}, gating~\cite{iwakiri_gate-defined_2022,iwakiri_tunable_2024}, or by employing strong magnetic fields to confine electrons along the one-dimensional edges of a quantum Hall droplet~\cite{klitzing_new_1980,ozyilmaz_electronic_2007,amet_selective_2014,nichele_edge_2016,nguyen_decoupling_2016,biswas_anomalous_2024}.

Recently~\cite{zhao_electron_2023}, we proposed a protocol for directing electrons using inverted-band $pn$-junctions, without relying on external spatial confinement of the electron trajectories. Such $pn$ junctions filter the electronic trajectories via a Klein tunneling-alike effect, allowing passage based on the incident angle at the $pn$ interface~\cite{zhao_electron_2023}. Unlike  Klein tunneling in Dirac materials~\cite{cheianov_selective_2006}, the filtering occurs at finite incident angles that are readily controlled by tuning the chemical potential of the junction. Crucially to this work, a narrow window of negative refraction scattering processes -- involving both particle-like and hole-like Fermi surfaces -- manifests and generates interfering closed paths for the electron trajectories. 

\begin{figure}[t!]
	\centering
	\includegraphics[scale=1]{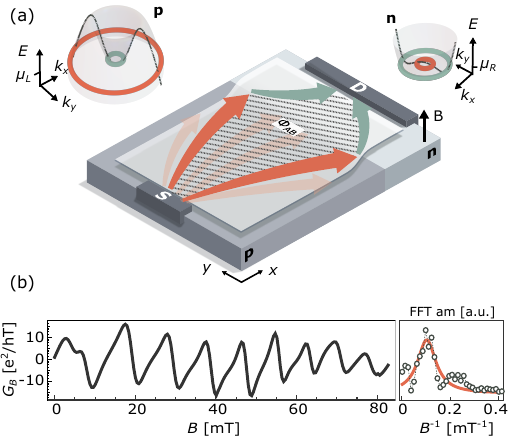}
    \caption{Device and Aharonov-Bohm oscillations. (a) A sketch of an inverted band $pn$ junction subject to an out-of-plane magnetic field $\mathbf{B}$. 
    Conductance is measured between the source lead (S) and the drain lead (D).
    Insets: sombrero-like bandstructure for both $p$- and $n$-doped regions with marked chemical potentials $\mu_{L/R}$, cf.~Eq.~\eqref{eq:H continuous}. Red (green) circles and arrows denote hole-like (electron-like) branches and trajectories, respectively. 
    (b) Left: Numerically-obtained magneto-differential conductance $G_B = \dd G/ \dd B$, cf.~Eq.~\eqref{eq: non-local conductance}. Right:  Fast Fourier transform of the pattern on the left (dots) with a fitted Lorentzian (orange line). In all figures of the paper, we use the following parameters:
    $\mathcal{A}=16\,\mathrm{meV\,nm}$, $\mathcal{M}_0=-30\,\mathrm{meV}$, $\mathcal{M}_1=400\,\mathrm{meV\,nm}^2$ and $\mathcal{M}_2=700\,\mathrm{meV\,nm}^2$. The $p$ and $n$-doped regions are achieved by setting $\mu_L=-28$ meV and $\mu_R=33$ meV. 
    }
    \label{fig: setup and result1}
\end{figure}

In this work, we propose how to realize an Aharonov-Bohm interferometer using a two-dimensional inverted-band $pn$ junction. The AB effect in our setup arises from two key ingredients: (i) negative refraction between distinct Fermi surfaces in the $p$- and $n$-doped regions, which creates closed loops of different electronic trajectories between point-like source and drain leads, and (ii) angular filtering of trajectories at the $pn$ interface, resulting in a limited number of trajectories that coherently interfere upon reaching the drain lead. In the presence of a weak perpendicular magnetic field, we predict that clear AB oscillations appear due to nonlocal interference. We numerically demonstrate the effect for a realistic experimental setting by adapting the band structure of InAs/GaAs~\cite{karalic_electron-hole_2020}. We further show that these oscillations persist even in the situation when the $pn$ interface has a finite length. Interestingly, we find that the frequency of the AB oscillations changes abruptly beyond a certain interface length, which we attribute to a shift in the maximal transmission angle due to a transition of the interface to an opaque barrier.
Our work brings magnetic control to electron optics applications in two-dimensional mesoscopic devices.

\textit{Model ---} To model the inverted band $pn$ junction (see Fig.~\ref{fig: setup and result1}), we consider the following 2D spinless Hamiltonian~\cite{karalic_electron-hole_2020,zhao_electron_2023},  
\begin{multline}
    \label{eq:H continuous}
    \mathcal{H}_{L/R}=(\mathcal{M}_0+\mathcal{M}_2\mathbf{k}^2)\sigma_z\\+(\mathcal{M}_1\mathbf{k}^2-\mu_{L/R}) \, \mathbb{1}_{2\times2} + \mathcal{A}\sigma_x \, ,
\end{multline}    
where $\sigma_i$ are Pauli matrices acting on two bands, and $\mathbf{k}$ denotes a 2D wavevector. The bare, uncoupled energy dispersions are particle- and hole-like in nature with respective effective masses $\mathcal{M}_1\pm\mathcal{M}_2$, where we assume  $\mathcal{M}_2>\mathcal{M}_1$. Particle-hole symmetry is maintained for $\mathcal{M}_1=0$. Such bare energy dispersions overlap with magnitude $\mathcal{M}_0$ and hybridize with coupling strength $\mathcal{A}$. The local chemical potential $\mu_{L/R}$, experimentally tunable by backgates~\cite{karalic_electron-hole_2020}, takes different values in the $p$-doped and $n$-doped regions. We assume these regions to respectively reside on the left- and right-hand sides of the $pn$ interface, i.e., $\mu_{L}<0$ and $\mu_R>0$. Diagonalizing the Hamiltonian~\eqref{eq:H continuous} yields the two dispersions $E(\mathbf{k})=\mathcal{M}_1\mathbf{k}^2-\mu_{L/R}\pm\sqrt{(\mathcal{M}_0+\mathcal{M}_2|\mathbf{k}|^2)^2+\mathcal{A}^2}$.
In the inverted-band regime, when $\mathcal{M}_0<0$, the resulting band structure resembles two sombreros facing one another, and separated by a gap, see Fig.~\ref{fig: setup and result1}(a). 
We label the spinor eigenstates of~\eqref{eq:H continuous} as $\psi^{\mathrm{L}/\mathrm{R},\pm}_{p/h}$, where $\mathrm{L}/\mathrm{R}$ refers to the left- and right-hand side of the junction and $p/h$ indicates the particle or hole-like Fermi surface, cf. Fig.~\ref{fig: setup and result1}(a). The superscript $\pm$ denotes the direction of group velocity relative to the $pn$ interface.
In the following, we consider the $pn$ junction when the chemical potentials, $\mu_{L/R}$, lie within the band-overlap energy interval of each sombrero. 
Specifically, we are interested in the response of the junction to an external perpendicular magnetic field, $\mathbf{B}= B \hat{z}$.

\textit{Sharp interface setup---} To numerically probe the transport properties of the system~\eqref{eq:H continuous}, we incorporate the aforementioned junction into the numerical toolbox Kwant~\cite{groth_kwant_2014}. To simulate a realistic transport experiment,  
we draw the junction along the $x$-direction, and include two metallic leads on the junction's ends. These source (S) and drain (D) leads are positions on the left and right sides, respectively, and have finite widths in the $y$-direction, which are much shorter than the sample size, see Fig.~\ref{fig: setup and result1}(a). The spatial distance between the junction interface and the leads is chosen carefully, such that injected hole-like states from S maximally refocus in D as particle-like states~\cite{supmat}. Similarly, the spatial extent of the junction in the $y$-direction is chosen such that the refocusing scattering states will not hit the boundaries. To remove background noise from finite-size effects, we attach additional absorbing leads along the top and bottom boundaries. The system's dimensions we use are taken relative to realistic Fermi wavelengths of InAs/GASb inverted-band semiconductors~\cite{karalic_electron-hole_2020,supmat}. The perpendicular magnetic field $\mathbf{B}$ is introduced using standard Peierls substitution~\cite{peierls_zur_1933}. We then study the two-terminal conductance between the leads by numerically evaluating the Landauer-Buttiker formula~\cite{buttiker_absence_1988}
\begin{equation}    \label{eq: non-local conductance}
    G=\frac{e^2}{h}\sum_{n=1}^{N}\sum_{m=1}^{M}\mathrm{Tr}[t^{\dagger}_{nm}t_{nm}],
\end{equation}
where $t_{nm}$ is the transmission coefficient from the $n$-th mode in the source lead to the $m$-th mode in the drain lead. The total numbers of modes $N$ and $M$ in the leads are determined by the widths of the leads.

\textit{Main result---}
In Fig.~\ref{fig: setup and result1}(b), we plot the resulting two-terminal conductance for magnetic fields ranging between $0 \sim 80$mT. We find that the magneto-differential conductance $G_B \equiv \dd G/\dd B$ exhibits clear periodic oscillations with an average period $\Delta \bar{B}\approx 8.43\,\mathrm{mT}$. As we show below, such oscillations in $G_B$ arise due to a mesoscopic AB effect whose origin lies in negative refraction at the interface. This is the main result of our work. 
We emphasize that such an emergence of AB oscillations in a simple 2D $pn$ junction that does not rely on any external spatial confinement is surprising~\cite{ihn_semiconductor_2009}.


\begin{figure}[t!]
	\centering
	\includegraphics{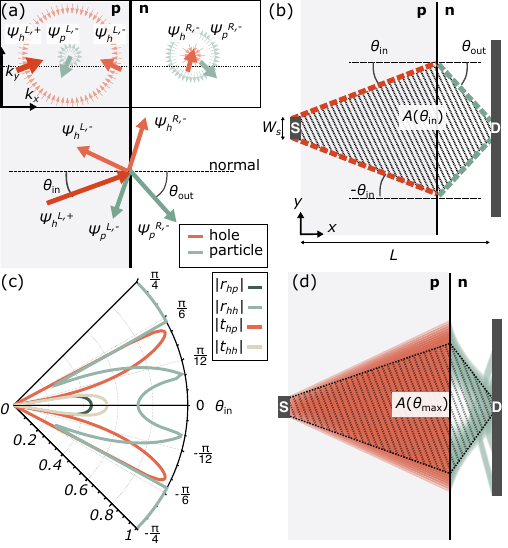}
    \caption{
    Scattering processes, negative refraction, and interfering paths. 
    (a) Upper panel: In reciprocal space, the scattering states and corresponding angles for each $k_y$ are determined by their location on the Fermi surface. Lower panel: In real space, the scattering angles are defined with respect to the normal (dashed line) of the $pn$ interface (solid black line). The angles $\theta_{\text{in}}$ and $\theta_{\text{out}}$ mark the incident angle of a hole-like state and the outgoing angle of a transmitted particle-like state.
    (b) Two trajectories of a hole-to-particle transmission scattering process with $\pm \theta_{\text{in}}$ that enclose a finite shaded area $A(\theta_{\rm in })$, cf.~Eq.~\eqref{eq: enclosed area}. 
    (c) Scattering amplitudes as a function of the incident angle for all processes involving an incident hole-like state. 
    (d) Trajectories of the incident hole-like state injected from the source lead, with their opacity determined by the scattering amplitudes in (c). 
    The maximal transmitted angle $\theta_{\text{max}}$  and the finite width of $S$ define the enclosed $A(\theta_{\text{max}})$ between the source and drain lead. Here we omit trajectories of the transmitted hole-like state in the $n$-doped region for clarity.
    }
    \label{fig: scattering and phase}
\end{figure}

To explain the mechanism of the oscillations in our system, we first notice that, in our regime of weak magnetic fields, the trajectories of the propagating states are only weakly affected by the magnetic field, i.e., they remain describable by ``ray tracing'' that emanate from the source in the $p$- and $n$-doped regions, separately~\cite{supmat}. 
Hence, in this limit, AB oscillations can emerge only when the following two prerequisites are satisfied: (i) the interface between the $p$- and $n$-doped region facilitates negative refraction that refocuses the trajectories at the drain lead where they interfere, and (ii) only a narrow distribution of incident angles participates in the negative refraction, such that the interference pattern survives decoherence via averaging over the contributions of many incident trajectories that carry different phases. Shown below, the latter will be the result of filtering by the transmission coefficients at the $pn$ interface.

To demonstrate (i), we consider an incident hole-like spinor  $\psi^{L,+}_{h}$ from the source with momentum $\mathbf{k}=(k_x,k_y)$, see Fig.~\ref{fig: scattering and phase}(a). 
We assume that the $pn$ junction is homogeneous along the $y$-direction, such that $k_y$ is conserved in all scattering processes. Thus, we can determine the outgoing states $\psi^{L/R,-}_{h/p}$ and their corresponding momenta $(k_{p/h}^{L/R,-},k_y)$ from momentum conservation at the Fermi surface, cf.~band structures in insets of Fig.~\ref{fig: scattering and phase}(a), and Ref.~\cite{zhao_electron_2023}. 
Note that the incident and outgoing states are defined with respect to the $pn$ interface, as illustrated in Fig.~\ref{fig: scattering and phase}(a).
The propagating direction of each scattering state is given by its group velocity $\mathbf{v}=\nabla_{\mathbf{k}}E(\mathbf{k})|_{\mathbf{k}=(k_{p/h}^{L/R,-},k_y)}$, see arrows in upper panel of Fig.~\ref{fig: scattering and phase}(a). Crucially, the hole-to-particle transmission scattering process exhibits negative refraction at the interface, i.e., for $\psi^{L,+}_{h}$ with a positive incident angle $\theta_{\mathrm{in}}=\sin^{-1}(k_y/|\mathbf{k}|)>0$, the outgoing particle-like spinor $\psi^{R,-}_{p}$ in the $n$-doped region propagates at a negative angle $\theta_{\mathrm{out}}<0$. This trajectory can interfere at the drain lead with another trajectory where $\psi^{L,+}_{h}$ impinges on the $pn$ interface at $-\theta_\mathrm{in}$, see Fig.~\ref{fig: scattering and phase}(b). Crucially, the two interfering paths encircle a kite-shaped area
\begin{align}   \label{eq: enclosed area}
    A(\theta_{\rm in})=L^2\tan(\theta_{\rm in})\frac{R_{pn}}{R_{pn}+1} + A_\mathrm{lead} \, ,
\end{align} 
where $L$ is the distance between S and D. The value of $R_{pn}$  represents the ratio between the length of the $p$- and $n$-doped region, which is chosen according to the focal point of trajectories~\cite{supmat}, see also Fig.~\ref{fig: scattering and phase}(b). 
Finally, $A_\mathrm{lead}= L W_s$ accounts for the added area introduced by the finite width $W_s$ of the source lead. 

To check if (ii) is satisfied, we employ a scattering matrix formalism~\cite{supmat} to calculate the transmission $t_{hp}$, $t_{hh}$ and reflection $r_{hp}$ and $r_{hh}$ scattering coefficients as a function of the incident angle, $\theta_{\rm in}$, where we assume an incident hole-like spinor (first subscript = $h$) and consider all four possible outgoing cases, see Fig.~\ref{fig: scattering and phase}(c). At most angles the dominant process is a reflection of a hole, $r_{hh}$, making the junction very opaque. Interestingly, we find that $t_{hp}$, which leads to negative refraction, maximizes in a narrow range around a finite incident angle $|\theta_{\text{in}}|\approx \theta_{\text{max}}= 5\pi/36$~\cite{supmat}. In other words, electrons transmitted through the $pn$ junction are mostly carried by trajectories incident around $\theta_{\mathrm{max}}$, while the other angles are screened by the $pn$ interface. 
We demonstrate this effect by drawing the ray trajectories of the hole-like states injected from the source and transmitted as particle-like states into the drain lead, see Fig.~\ref{fig: scattering and phase}(d). The drawn trajectories' opacity is taken according to their transmission amplitude at the interface, cf. Fig.~\ref{fig: scattering and phase}(c). Thus, we find a large depleted area due to the transmission filtering at small angles and high density only in a narrow scattering angle region.
Note that by taking only the two trajectories with the maximal $t_{hp}$ into account, defined by $\theta = \pm \theta_{\mathrm{max}}$ and marked by dashed lines in Fig.~\ref{fig: scattering and phase}(d), we can use Eq.~\eqref{eq: enclosed area} to obtain $A(\theta_{\rm max})=4.69\times 10^5\ \text{nm}^2$ for $W_s=80\,\mathrm{nm}$ and $R_{pn}=131/44$ in our setup.

\begin{figure*}[t]
		\centering
		\includegraphics[scale=1]{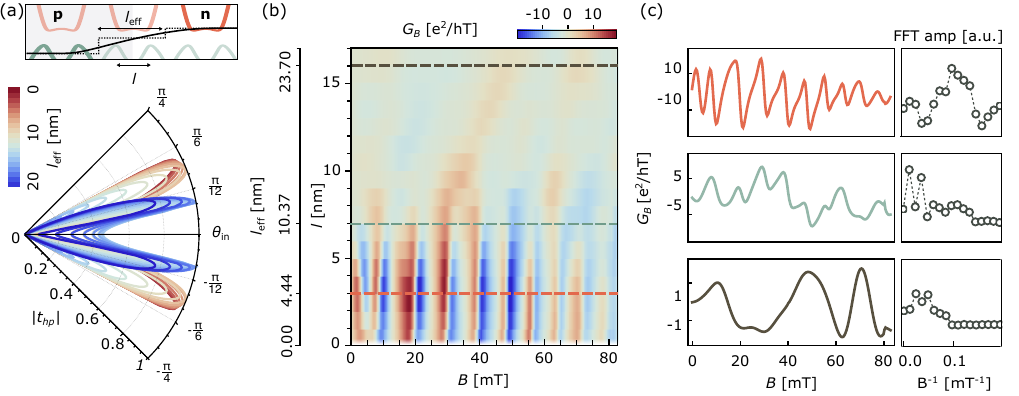}
    	\caption{Smooth interface. 
        (a) Upper panel: A spatially varied chemical potential (solid line) induces an effective barrier defined by the length over which the Fermi level is inside the gap, leading to an effective two-interface junction (dotted line). 
        Lower panel: Transmission amplitude of a hole-to-particle  $|t_{hp}|$ scattering as a function of the incident angle $\theta_{\rm in}$ and the effective barrier length $l_\mathrm{eff}$.
        (b) Numerically obtained $G_B$ as a function of the magnetic field $B$ and the (effective) barrier length ($l_\mathrm{eff}$) $l$. (c) Left: Cuts of (b) at $l=3,\ 7,\ 16\, \text{nm}$ (red, green, black), cf.~dashed lines. Right: Fast Fourier transform of the patterns. 
        }
    \label{fig: barrier width and robustness}
\end{figure*}

Each scattering trajectory follows a path $\mathcal{P}$ and accumulates a dynamical phase $\phi_D=\exp(\int_{\mathbf{x}\in\mathcal{P}}\mathrm{d}\mathbf{l}\cdot\mathbf{k})$. To these paths, we can introduce the perpendicular magnetic field as a vector potential $\mathbf{A}=(-By/2,Bx/2,0)$, where it may impact the scattering amplitudes in three ways: (i) bend the trajectories due to a Lorentz force, (ii) alter the scattering angles and amplitudes at the junction, and (iii) introduce an Aharonov-Bohm (AB) phase to the interfering paths via minimal coupling, $\exp(\int_{\mathbf{x}\in\mathcal{P}}\mathrm{d}\mathbf{l}\cdot\mathbf{A})$. Neglecting (i) and (ii) due to the weak magnetic fields we used, we can divide the sum in Eq.~\eqref{eq: non-local conductance} into terms that do not include interfering paths (where the accumulated phase is irrelevant) and interfering scattering trajectories that are impacted by the magnetic flux through the area A. We can, moreover, assume that the dominant magnetic-field dependent scattering process involves the paths at $\theta_{\rm in}=\pm\theta_{\mathrm{max}}$~\cite{supmat}, leading to
\begin{align}    \label{eq: AB oscillation}
    G_B
    \approx \frac{e^2}{h}
    |t_{hp} (\theta_{\rm max})|^2  
    \frac{\dd}{\dd B} \Big[ 1+\cos(\phi_{\mathrm{tot}})
    \Big] \, ,
\end{align}
where $\phi_{\mathrm{tot}}=eB A(\theta_{\rm max})/h$ is the flux threading the enclosed area $A(\theta_{\rm max})$. 
The magnetic response~\eqref{eq: AB oscillation}, therefore, exhibits a period $\Delta B=h  \left[e A(\theta_{\rm max})\right]^{-1}$ that is solely determined by the AB phase. In our case, we obtain a period of $\Delta B = 8.82$ mT, which is in good agreement with the one extracted from the numerics, cf.~Fig.~\ref{fig: setup and result1}(b).

\textit{Graded $pn$ interface---}
Perfectly sharp $pn$ interfaces cannot be realized in experiments. We, therefore, proceed to check how the predicted AB oscillations behave in the presence of a graded interface. We model it by introducing a spatially modulated chemical potential $\mu(x)=\mu_R+(\mu_L-\mu_R)[1-\tanh(x/l)]/2$, where $l$ is its characteristic length, see in Fig.~\ref{fig: barrier width and robustness}(a). Importantly, this interpolation introduces an effective gapped region between the $p$ and $n$ regions. We numerically calculate $G_B$ as a function of $l$ and show it in Fig.~\ref{fig: barrier width and robustness}(b). 
We observe a clear change in the oscillating pattern at $l \approx 7\,\text{nm}$.
More quantitatively, the frequency of oscillation in $G_B$ reduces for smoother interfaces, cf.~three cuts at $l=3\,\text{nm}$, $7\,\text{nm}$, and $16\,\text{nm}$ in Fig.~\ref{fig: barrier width and robustness}(c). 

To gain a deeper understanding of the observations discussed above, we analytically describe an effective potential barrier induced by the smooth interface by employing an effective two-interface model.
As the Fermi level leaves the upper/lower band at energies $\Lambda_{\pm}=(-\mathcal{M}_1\mathcal{M}_0\pm A\sqrt{\mathcal{M}_2^2-\mathcal{M}_1^2})/\mathcal{M}_2$, the barrier between the $p$- and $n$-doped regions has an effective length $l_\text{eff}=x_+-x_-$with $x_{\pm}=\mathrm{arctanh}(1-2(\Lambda_{\pm}-\mu_R)/(\mu_L-\mu_R))l$. In this simplified model, we assume the interfaces between the three regions to be sharp; see the dashed line in the upper panel of Fig.~\ref{fig: barrier width and robustness}(a).
Using the S-matrix formalism, we once more calculate the transmission amplitude $t_{hp}$ as a function of the incident angle and effective barrier length $l_\mathrm{eff}$~\cite{supmat} and show the result in the lower panel of Fig.~\ref{fig: barrier width and robustness}(a). For a short barrier $l_\mathrm{eff}< 10\ \text{nm}$, we obtain a similar behavior as in the previous case of a vanishing barrier, with negative refraction centered around $\theta_{\text{max}}\approx5\pi/36$. Interestingly, with increasing $l_\mathrm{eff}$, we find a crossover in the behavior of $|t_{hp}|$, i.e., the maximal transmitted angle $\theta_{\rm max}$ undergoes an abrupt transition at $l_\mathrm{eff}\approx 10\ \text{nm}$. As a result, for a longer interface $l_\mathrm{eff}> 10\ \text{nm}$, $|t_{hp}|$ maximizes at a smaller angle of $\theta_{\rm max}' \approx \pi/18$, which implies a smaller enclosed area $A(\theta_{\rm max}')$, and a different AB-oscillation period that follows from Eq.~\eqref{eq: AB oscillation}, which is in agreement with the numerical results, cf.~Fig.~\ref{fig: barrier width and robustness}.
We attribute this transition to a competition between the length of the barrier and the decay length of evanescent waves penetrating the barrier from the interfaces~\cite{yao_anomalous_2024}:  the system enters the tunneling limit when the electrons only tunnel through the overlapping evanescent tails, resulting in an abrupt change of the $\theta_{\rm max}$~\cite{supmat}. 


In conclusion, we propose the realization of an AB-interferometer using negative refraction in a two-dimensional inverted-band $pn$ junction. A key difference, compared with common AB interferometers, is that we do not impose external spatial confinement on the trajectories of electrons; instead, the confinement emerges naturally due to the band structure's characteristic sombrero shape. Specifically, using the S-matrix formalism, we find that the negative refraction transmission amplitudes are facilitated by hole-to-particle scattering $t_{hp}$, whose distribution maximizes in a narrow range of incident angles. The latter realizes an effective spatial confinement that guides electrons~\cite{zhao_electron_2023}. This emergent spatial confinement entails that all transmitted electronic trajectories exhibit a kite-shaped enclosed surface which, under the external magnetic field, gives rise to unique AB oscillations in conductance. Commonly, one would expect a reduction in the transmission due to a smooth interface~\cite{cheianov_selective_2006}, while, on the contrary, we find that the oscillations persist with a distinct change in the AB-oscillation period.
Moreover, we tested the robustness of the AB oscillations against finite temperature and disorder~\cite{supmat}. We observe clear oscillations up to temperatures of $1K$ and disorder strength of several meV-s, which gives strong indications that the proposed setup can be realized in realistic experimental situations.
The proposed mechanism of generating AB interference is related to the sombrero hat shape of the band structure, which can be found in a plethora of systems; including $s$-wave and chiral $p$-type superconductors~\cite{strkalj_tomasch_2024}, heterostructures with strong Rashba spin-orbit interaction like HgTe/CdTe~\cite{minkov_two-dimensional_2013} and InAs/GaSb~\cite{karalic_electron-hole_2020}, two-dimensional monochalcogenides like GaSe~\cite{li_controlled_2014} and transition metal dichalcogenides like WTe$_2$~\cite{qian_quantum_2014}.
Moreover, the Bernal bilayer graphene, which is known for its tunability of the sombrero-hat bandstructure under a biased voltage~\cite{min_ab_2007,castro_biased_2007,mccann_low_2007}, suggests potential realizations of AB interferometer with graphene-based material. 
With all of the aforementioned, we think that the physics discussed in this work can be experimentally tested in current state-of-the-art mesoscopic platforms.

\section*{acknowledgments}
This work was supported by the Deutsche Forschungsgemeinschaft (DFG) through project number 449653034, and ETH research grant ETH-28 23-1. The work of A.\v{S}. is supported by the European Union’s Horizon Europe research and innovation programme under the Marie Sk\l{}odowska-Curie Actions Grant agreement No. 101104378. 
%

\newpage
\cleardoublepage
\setcounter{figure}{0}

{\onecolumngrid
\begin{center}
	\textbf{\normalsize Supplemental Material: Aharonov-Bohm interferometer in inverted-band pn junctions}\\
	\vspace{3mm}
	\vspace{4mm}
	
	
\end{center}}
\twocolumngrid
\setcounter{secnumdepth}{2}
\setcounter{equation}{0}
\setcounter{section}{0}
\setcounter{figure}{0}
\setcounter{table}{0}
\setcounter{page}{1}
\makeatletter
\renewcommand{\bibnumfmt}[1]{[#1]}
\renewcommand{\citenumfont}[1]{#1}
\renewcommand{\theHfigure}{Supplement.\thefigure}

\setcounter{enumi}{1}
\renewcommand{\theequation}{S\arabic{equation}}
\renewcommand{\thesection}{S\arabic{section}}
\renewcommand{\thefigure}{S\arabic{figure}}
\setcounter{section}{0}
\renewcommand{\thesubsection}{S\arabic{section}.\arabic{subsection}}
\renewcommand{\thetable}{S\arabic{section}.\arabic{table}}

\section{Details on the numerical simulation using Kwant  \label{app:details_numerical_simulation}}
Throughout the paper, unless specified otherwise, we use the following parameters: $\mathcal{A}=16\,\mathrm{meV\,nm}$, $\mathcal{M}_0=-30\,\mathrm{meV}$, $\mathcal{M}_1=400\,\mathrm{meV\,nm}^2$ and $\mathcal{M}_2=700\,\mathrm{meV\,nm}^2$, which approximate the bandstructure of InAs/GaAs~\cite{karalic_electron-hole_2020}  [cf.~Eq.~\eqref{eq:H continuous} in the main text]. To numerically calculate the transport through the system described in the main text, we first discretize the continuous Hamiltonian and define a tight-binding Hamiltonian over a finite-size region containing $L=1050$ sites in the $x$-direction (length) and $W=1200$ sites in the $y$-direction (width). The length of the $p$- and $n$-doped region is $786\,\mathrm{nm}$ and $264\,\mathrm{nm}$, respectively. We set the lattice spacing to $1\,\mathrm{nm}$, such that  
for the aforementioned values of parameters, the smallest Fermi wavelength is about 15 times larger than the lattice spacing. This ensures that our lattice model reliably describes the continuum system at the given low energies energies that we use.  

We attach semi-infinite leads on each side of the junction, where the source lead width is $80$ sites and the drain lead width is $800$ sites. We assume that both leads are normal metals with parabolic dispersions, i.e., we adopt the Hamiltonian from Eq.~\eqref{eq:H continuous} and set $\mathcal{M}_0=30\,\mathrm{meV}$, $\mu_{L/R}=-10\mathcal{M}_0$ and $\mathcal{A}=0$ for that purpose. 
With the setup defined, we use Kwant~\cite{groth_kwant_2014} to obtain the non-local conductance between the source and the drain lead, corresponding to Eq.~\eqref{eq: non-local conductance} in the main text.

\section{Conductance at finite temperatures and in the presence of disorder}
In this section, we study the performance of the AB interferometer at finite temperatures and its robustness against disorder. 
First, to study the influence of finite temperatures: we calculate the temperature-dependent conductance as
\begin{equation}
    \langle G\rangle (T)=\frac{1}{2k_BT}\int^{\infty}_{-\infty}\mathrm{d}\varepsilon \, G(\varepsilon) \frac{1}{\cosh(\varepsilon/k_BT)+1} \, ,
\end{equation}
where $k_B$ is the Boltzmann constant and $G(\varepsilon)$ is conductance defined in Eq.~\eqref{eq: non-local conductance} in the main text at energy $\varepsilon$. We repeat the simulation $30$ times for different $\varepsilon$ sampled out of the distribution $P(\varepsilon)=[2 k_B T (\cosh(\varepsilon/k_BT)+1)]^{-1}$, such that the integral above is evaluated using a Riemann sum.

\begin{figure}[t!]
		\centering
		\includegraphics[scale=1]{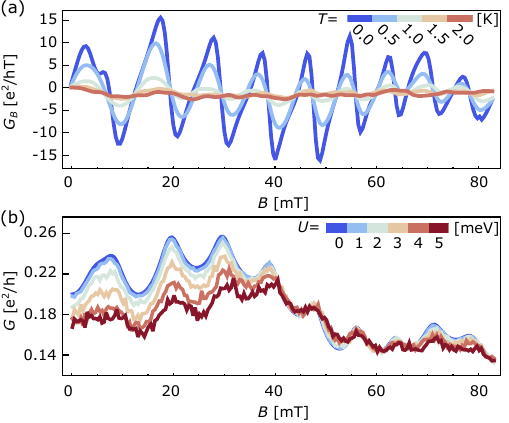}
    	\caption{Impact of finite temperature and disorder.
        (a) Magneto-differential conductance as a function of the magnetic field $B$ for different temperatures varying from $0\,\text{K}$ to $2\,\text{K}$. (b) Conductance as a function of $B$ for different disorder strengths. The disorder is introduced as an onsite potential fluctuation obeying a normal distribution with a variance $U$.}
    \label{fig: temperature and disorder}
\end{figure}

We show the results in Fig.~\ref{fig: temperature and disorder}(a), where in order to emphasize the oscillations, we plot $G_B(T)=\dd\langle G\rangle(T)/\dd B$  as a function of $B$. The amplitude of the oscillation decreases with increasing temperature and vanishes when the temperature approaches $T\sim2\,\mathrm{K}$. Nevertheless, for $T \leq 1\,\mathrm{K}$, the oscillation pattern is clearly visible, meaning that it should be possible to observe it in a realistic experimental situation. 

Similarly, we find that the oscillation pattern persists also when introducing moderate disorder. For the latter, we introduce a random onsite potential $\mu_{dis}$ sampled out of a normal distribution with zero mean and the standard deviation $U$. As shown in Fig.~\ref{fig: temperature and disorder}(b), the disorder introduces random fluctuations into the conductance $G$. Nonetheless, the oscillation pattern remains recognizable up to $U=5\,\mathrm{meV}$, which corresponds to a thermal fluctuation at $T_{dis}=U/k_B\approx58\,\mathrm{K}$.

\section{Scattering matrix formalism}
\label{app:scattering matrix approach}
We employ a scattering matrix (S-matrix) formalism~\cite{karalic_electron-hole_2020,zhao_electron_2023}
\begin{equation}    \label{eq:scattering equation1}
    \begin{pmatrix}
    \vec{\Phi}^{L,-}\\
    \vec{\Phi}^{R,-}
    \end{pmatrix}
    =\mathbf{S}
    \begin{pmatrix}
    \vec{\Phi}^{L,+}\\
    \vec{\Phi}^{R,+}
    \end{pmatrix} \, ,
\end{equation}
where the scattering amplitudes are encoded in the S-matrix $\mathbf{S}$ and $\vec{\Phi}^{L/R,\pm} = (\psi_p^{L/R,\pm}, \psi_h^{L/R,\pm})^T$ are spinors capturing the contributions of the particle- and hole-like branches, $p$ and $h$, located on the left (L) and right (R) side of the $pn$ interface. Furthermore, we denote incoming and outgoing states, with respect to the interface, with $+$ and $-$, respectively. The scattering matrix is given by
\begin{equation}\label{eq:s matrix}
    \mathbf{S}=\begin{pmatrix}
    \mathbf{r} & \mathbf{t}'\\
    \mathbf{t} & \mathbf{r}'
    \end{pmatrix}\,,
\end{equation}
with the reflection scattering amplitudes $\mathbf{r}$, $\mathbf{r}'$, and transmission scattering amplitudes $\mathbf{t}$ and $\mathbf{t}'$ being $2\times2$ matrices 
\begin{align}\label{eq:scattering amplitudes}
        \mathbf{r}&=\begin{pmatrix}
        r_{pp} & r_{ph}\\
        r_{hp} & r_{hh}
        \end{pmatrix}, &
        \mathbf{t}&=\begin{pmatrix}
        t_{pp} & t_{ph}\\
        t_{hp} & t_{hh}
        \end{pmatrix},\nonumber\\
        \mathbf{r}'&=\begin{pmatrix}
        r'_{pp} & r'_{ph}\\
        r'_{hp} & r'_{hh}
        \end{pmatrix}, &
        \mathbf{t}'&=\begin{pmatrix}
        t'_{pp} & t'_{ph}\\
        t'_{hp} & t'_{hh}
        \end{pmatrix},
\end{align}
where $t_{a, a'}$ and $r_{a, a'}$ indicate the amplitude of the transmission/reflection from a $a$-like spinor ($a=p,h$) to a $a'$-like spinor ($a'=p,h$) at the interface, respectively. 
We impose the continuity condition for the total wavefunctions $\Psi^{L/R}(\mathbf{x},k)$ and their first derivatives at the interface, i.e. at $x=0$, obtaining the set of equations
\begin{equation}\label{eq: continuity condition}
    \begin{split}
        \Psi^{L}(\mathbf{x},k)\big|_{x=0}&= \Psi^{R}(\mathbf{x},k)\big|_{x=0} \, ,\\
        \partial_x\Psi^{L}(\mathbf{x},k)\big|_{x=0}&= \partial_x\Psi^{R}(\mathbf{x},k)\big|_{x=0} \, .
    \end{split}
\end{equation}

As an example, for the scattering process $t_{hp}$ discussed in the main text, we start with a hole-like spinor with $k_y=k$ impinging from the left. The wavefunctions in the $p$-doped region on the left- and $n$-doped region on the right-hand side are then given by
\begin{align}   \label{eq: psiL R}
   &\resizebox{\columnwidth}{!}{$\displaystyle{ \Psi^{L}(\mathbf{x},k)=\psi^{L,+}_{h}e^{-i\mathbf{k}_{h}^{L,+}\mathbf{x}}+\tilde{r}_{hp}\psi^{L,-}_{p}e^{-i\mathbf{k}_{p}^{L,-}\mathbf{x}}+\tilde{r}_{hh}\psi^{L,-}_{h}e^{-i\mathbf{k}_{h}^{L,-}\mathbf{x}} \, , \nonumber}$}\\
    &\Psi^{R}(\mathbf{x},k)=\tilde{t}_{hp}\psi^{R,-}_{p}e^{-i\mathbf{k}_{p}^{R,-}\mathbf{x}}+\tilde{t}_{hh}\psi^{R,-}_{h}e^{-i\mathbf{k}_{h}^{R,-}\mathbf{x}} \, ,
\end{align}
where $\mathbf{k}^{L/R,\pm}_{p/h}=(k^{L/R,\pm}_{p/h},k)^\mathrm{T}$ is the momentum of spinors with $k_y=k$ and $k_x=k^{L/R,\pm}_{p/h}$.
Using Eqs.~\eqref{eq: psiL R} in Eqs.~\eqref{eq: continuity condition}, we can determine the amplitudes of $\tilde{t}_{hp}$ as well as $\tilde{t}_{hh}$, $\tilde{r}_{hp}$ and $\tilde{r}_{hh}$. 
Lastly, to respect the conservation of the spinor current in the direction of the interface normal, the obtained amplitudes are modified by the ratio of group velocities of the initial and final states in the direction of the interface normal, i.e., $x$-direction~\cite{ihn_semiconductor_2009}. For instance, the normalized scattering amplitude $t_{hp}$ is given by
\begin{equation}
t_{hp}=\sqrt{\frac{v_p^R}{v_h^L}} \, \tilde{t}_{hp}\,,
\end{equation}
where $v_{p/h}^{L/R}=|\partial_x E(\mathbf{k})|\big|_{\mathbf{k}=\mathbf{k}^{L/R,\pm}_{p/h}}$ is the $x$-component of the group velocity for the corresponding spinor $\psi^{L/R,\pm}_{p/h}$. Note that for the bound states that might appear at the interface, i.e. the states with $\textrm{Im}(k_x)\neq0$, the group velocity is vanishing $[v_{p/h}^{L/R}]_x=0$.

\begin{figure}[!t]
	\centering
	\includegraphics{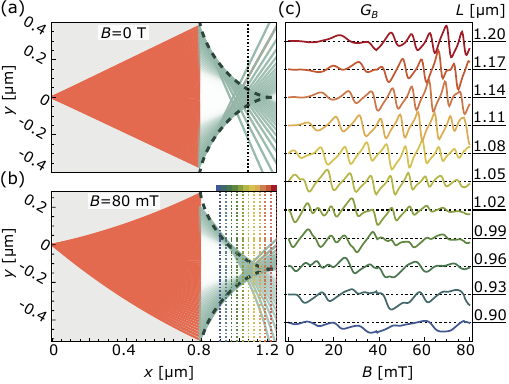}
    \caption{Changing the distance to the drain lead.
    (a) Trajectories of hole-like and particle-like spinors (red and green lines, respectively) for a vanishing magnetic field. In the $n$-doped region (on the right), trajectories of all transmitted particle-like states can be contoured by an envelop function $F(x)$, cf. Eq.~\eqref{eq: envelope}, marked by the dark green dashed line. For the results shown in the main text, we position the drain lead at $x=1.05\,\mathrm{\mu m}$ (black dotted line). 
    (b) Under a magnetic field of $B=80\,\mathrm{mT}$, the bent trajectories on the right remain well enveloped by the same function $F(x)$ up to an overall shift in the $y$-direction $\delta F=-0.13$, which is obtained as a fitting parameter for different magnetic fields.
    (c) The magneto-differential conductance $G_{B}$ for the $n$-doped region terminated at different positions $L$ from the source lead, i.e., the drain lead is positioned at different $x=L$ shown in (b), marked by dash lines and colored correspondingly.  
    }
    \label{fig: envelope and focal point}
\end{figure}

\section{The effective path and the focal point \label{app:effective_path}}

To numerically illustrate the electronic trajectories injected from the source lead as well as the scattering at the interface, we calculate their local density of the polarization (LDOP)
\begin{equation}
\mathrm{LDOP}=\sum_{n=1}^{N}\Psi^\dagger_n(\mathbf{x})\sigma_z\Psi_n(\mathbf{x}) \, ,
\end{equation}
where $\sigma_z$ is a Pauli matrix and $\Psi_n(\mathbf{x})$ is the wavefuncion of the $n$-th outgoing mode from the source lead. Note that the discrete modes $n$ originate from the finite size (in the $y$-direction) of the source lead.

\subsection{Effective path as the envelope of trajectories}
As explained in the main text, the effective spatial confinement for the electrons is realized by the fact that the transmission amplitude $|t_{hp}|$ has a finite value within a narrow range of incident angles around $\theta_{\mathrm{max}}$, see Fig.~\ref{fig: scattering and phase}(c) in the main text. 
In Fig.~\ref{fig: envelope and focal point}(a), we show all trajectories of transmitted particle-like spinors with their opacity weighted by the corresponding transmission amplitude $|t_{hp}|$. 
We find that the resultant effective path of the particle-like states from the $pn$ interface to the drain lead, where the trajectories are densely distributed, is a convex-shaped curve [see the dark green dashed line in Fig.~\ref{fig: envelope and focal point}(a)]. This curve is defined by the envelope of all trajectories in the $n$-doped region (on the right-hand side from the $pn$ junction). In other words, each point on this curve is tangent to a trajectory of the transmitted particle-like spinor. For the upper paths, the expression is hence given through the Legendre transformation:
\begin{equation}\label{eq: envelope}
    F(x)=qx-f(q),
\end{equation}
where $q$ is the slope of each trajectory. The function $f(q)$ is the corresponding $y$-intercept at the $pn$ interface
\begin{equation}
    f(q)=-\frac{qL_Lk_R}{k_L\sqrt{1+q^2}\sqrt{1-\frac{{k_R}^2q^2}{{k_L}^2(1+q^2)}}},
\end{equation}
where $L_L$ is the length (in the $x$-direction) of the $p$-doped region on the left-hand side and $k_{L/R}$ is the Fermi momentum of the hole/particle-like branch in the $p$/$n$-doped region, respectively. For the lower trajectories, i.e. the ones with negative incident angles, the above expression is simply $-F(x)$, reflecting the symmetry with respect to the $y=0$ axis. Moreover, for the magnetic field ranging from $0\sim80\,\mathrm{mT}$, $F(x)$ remains to be a good envelope for the bent trajectories, see in Fig.~\ref{fig: envelope and focal point}(b) up to a constant shift in the $y$-direction, i.e. $F(x)\rightarrow F(x)+\delta F$. The exact value of $\delta F$ can be obtained by fitting.

\subsection{Changing the distance to the drain lead}
We obtain the focal point $x_0$, measured from the $pn$ interface, of the effective path for particle-like states by solving $F(x)=0$ 
\begin{equation}
    x_0=L_L \left(1+\frac{k_R}{k_L} \right) \, .
\end{equation}
Using the Hamiltonian parameters from the main text, and setting $L_L=786\,\mathrm{nm}$, we obtain $x_0 \approx 1.158\,\mathrm{\mu m}$. As shown in Fig.~\ref{fig: envelope and focal point}, not all the trajectories perfectly focus on $x_0$ even in the absence of the magnetic field. This is a direct consequence of a broken particle-hole symmetry of the Hamiltonian~\eqref{eq:H continuous}.
Restoring the particle-hole symmetry by setting $M_1=0$ would give a perfect focusing of all trajectories in a single point~\cite{zhao_electron_2023} whose coordinates are given by $(x_0, y_0=0)$, see Fig.~\ref{fig: envelope and focal point}(a).

To investigate the effect of the position of the drain lead on the oscillations in $G_B$, we perform numerical simulations with samples having different lengths $L$, but with a fixed length of the $p$-doped region $L_L=786\,\mathrm{nm}$, which corresponds to varying position of the drain lead in the x-direction with respect to the $pn$ interface. 
The result is shown in Fig.~\ref{fig: envelope and focal point}(c). The oscillation pattern in $G_B$ is pronounced for a wide range of $L$, as long as $L$ is relatively close to $x_0$; clear periodic oscillations are visible in the range of $L$ from $1 \,\mathrm{\mu m}$ to $1.2\,\mathrm{\mu m}$. We hence choose $L=1.050\,\mathrm{\mu m}$, where the clear oscillation pattern is present in the whole range of probed $B$. Correspondingly, the ratio between the length of $p$- and $n$-doped region is then given by $R_{pn}=0.786/(1.050-0.786)=131/44$.

\section{Dynamical and Aharonov-Bohm phases \label{sec: dynamical and Aharonov-Bohm phases}}
In the following, as well as in the main text, the magnetic field is applied perpendicularly to the sample, i.e. in the $z$-direction, $\vec{B}=(0,0,B)^T$.

\subsection{Electronic trajectories under the magnetic field}
\begin{figure}[t!]
		\centering
		\includegraphics{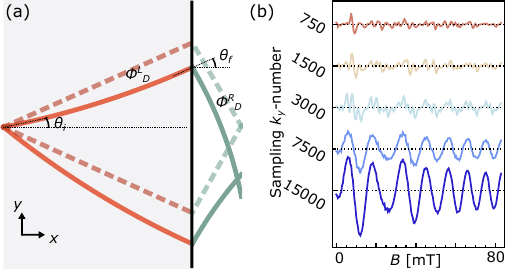}
		\caption{Evaluating the accumulated phase  in the magnetic field. (a) A sketch of the area enclosed by the trajectory of electrons propagating from the source to the drain under the magnetic field. The bent trajectory is characterized by its initial angle at the source $\theta_i$ and the final angle $\theta_f$ at the interface. The dashed lines denote  trajectories in the absence of the magnetic field. For a single path, a dynamical phase $\phi^{L/R}_D$ is accumulated along the trajectory.
        (b) $\dd |\Psi_t(B)|^2 /\dd B$ obtained by evaluating Eq.~\eqref{eq: probability density analytic} as a function of the magnetic field. The integral is evaluated using a Riemann sum with different numbers of sampling intervals for $k_y$; see the text for more details.}
    \label{fig: trajectory and analytic result}
\end{figure}

Subject to a finite perpendicular magnetic field, the trajectory of an electron propagating inside the system is bent by the Lorentz force and consequently samples a cyclotron orbit.  
This means that a state injected from the source lead with momentum $\mathbf{k}=(k_x,k_y)$ and an angle $\theta_{i}$ hits the $pn$ interface with a different angle after propagating over some distance between the lead and the $pn$ interface, see Fig.~\ref{fig: trajectory and analytic result}(a). In general, after the electron propagates over a distance $d$ in the $x$-direction, its angle --measured as the deflection from the normal of the $pn$ interface-- changes from $\theta_{i}$ to $\theta_{f}$ according to the formula:
\begin{equation}\label{eq:final angle}
    \theta_{f}=\arcsin\left(\pm\frac{d}{R}+\sin{\theta_i}\right),
\end{equation}
where $R=\hbar|\mathbf{k}|/(eB)$ is the cyclotron radius, and the sign $\pm$ corresponds to the hole- and particle-like spinor, respectively.

\subsection{Phase accumulation in the magnetic field \label{sec: phase accmulation}}
The dynamical phase accumulated as the band spinor is propagating along an arbitrary trajectory $\mathcal{P}$ located in the $p$- or $n$-doped region can be calculated as
\begin{equation}
\phi_D(\theta_i,R)=\int_{(x,y)\in\mathcal{P}}\mathbf{k}\cdot\mathrm{d}\vec{l}=|\mathbf{k}| \, R \, |\theta_f-\theta_i|
\end{equation}
where $\theta_i$ and $\theta_f$ are the initial and final angles of the line segment $\mathcal{P}$ being connected with Eq.~\eqref{eq:final angle}, and $R$ is the cyclotron radius. Furthermore, the field-induced phase accumulated along the same path is given as [using a symmetric gauge $\mathbf{A}=(-By/2, Bx/2, 0)^T$]
\begin{equation}
\begin{split}
    \phi_{B}(\theta_i,R)&=-\int_{(x,y)\in\mathcal{P}}\frac{e}{\hbar}\mathbf{A}\cdot\mathrm{d}\vec{l} \\
    &=-\frac{|\mathbf{k}|}{2}\Big[R\big(\delta \theta-\sin(\delta \theta)\big)-(x_i\delta y-y_i\delta x)\Big]
\end{split}
\end{equation}
where $(x_i,y_i)$ is the starting point of the trajectory and $\delta \theta \equiv \theta_f-\theta_i$. The displacement of the end point of the trajectory $\mathcal{P}$ compared with its beginning is denoted by $\delta x=R|\sin(\theta_f)-\sin(\theta_i)|$ and $\delta y=R|\cos(\theta_f)-\cos(\theta_i)|$. 

Next, we consider the part of the total wavefuncion at the drain lead that describes the transmission from a hole-like to a particle-like state as
\begin{equation}    \label{eq:wf_at_the_drain}
\begin{split}
    \Psi_t(B)=\int_{-\frac{\pi}{2}}^{\frac{\pi}{2}} \mathrm{d}\theta \,  t_{hp}(\theta) \, \psi_p(-\theta) \, e^{i\phi_t(\theta,B)} \, ,
\end{split}
\end{equation}
where $\theta$ is the incident angle of the hole-like spinor at the $pn$ interface, see Fig.~\ref{fig: trajectory and analytic result}(a) where this angle is marked by $\theta_f$. Here we use $\psi_p\equiv\psi^{R,-}_p$ for the particle-like spinor. Note that we take the boundaries of the integral over $\theta$ to be the same as for $\theta_i$, which is a valid approximation for the weak magnetic fields that we study in our setup, i.e. when $d=L_L \ll R(B)$.
At strong magnetic fields, some trajectories are bent strongly and never reach the $pn$ interface, changing the integral in Eq.~\eqref{eq:wf_at_the_drain}.

The total accumulated phase when a particle-like spinor reaches the drain lead is given by
\begin{align}   \label{eq:total phase}  \phi_t(\theta,B)=&\phi^{L}_D(\theta_i,B)+\phi^{L}_{B}(\theta_i,B) \nonumber\\
    &+\phi^{R}_D(-\theta,B)+\phi^{R}_{B}(-\theta,B),
\end{align}
where the superscript denotes the phase accumulated in the $p$- and $n$-doped region, on the left ($L$) and right ($R$) from the interface, respectively. The initial angle $\theta_i$ is obtained by solving Eq.~\eqref{eq:final angle} with $\theta_f=\theta$. We can then write the probability density of the total state at the drain lead as
\begin{equation}\label{eq: probability density analytic}
\begin{split}
    |\Psi_t(B)|^2=&\int_{-\frac{\pi}{2}}^{\frac{\pi}{2}}\int_{-\frac{\pi}{2}}^{\frac{\pi}{2}}\mathrm{d}\theta\mathrm{d}\theta't_{hp}^*(\theta')t_{hp}(\theta)\psi^{\dagger}_p(-\theta')\psi_p(-\theta)\\
    &\times e^{i(\phi_t(\theta,B)-\phi_t(\theta',B))}.
\end{split}
\end{equation}
We evaluate the above expression using a Riemann sum for discrete values of $\theta$.
On a technical note, the aforementioned values are obtained from evenly sampled $k_y$ momenta over the whole Fermi surface of holes on the left side of the $pn$ junction -- by using the expression $\theta = -\arcsin\left(k_y/\abs{\mathbf{k}}\right)$, we see that the angles in the whole interval $-\pi/2$ and $\pi/2$  are covered. 

In Fig.~\ref{fig: trajectory and analytic result}(b), we plot the evaluated $\partial |\Psi_t(B)|^2/\partial B$ and find that it is dominated by random fluctuations when the number of the sampling points of $k_y$ is small ($<7000$), while a clear pattern can be observed for a finer sampling ($\geq7500$). This is because the dynamical phase $\phi_D$ varies strongly for different trajectories, even if they are spatially close, whereas the field-induced phase $\phi_B$ varies slowly between different trajectories. As a result, the averaging over many different trajectories will give a vanishing contribution of the dynamical phase, and the probability density $|\Psi_t(B)|^2$ is mostly dominated by $\phi_{B}$ around $\theta_{\mathrm{max}}$. Therefore, omitting the $\phi_D$ terms in Eq.~\eqref{eq:total phase} and inserting the remain terms to Eq.~\eqref{eq: probability density analytic}, we obtain
\begin{widetext}
\begin{align}\label{eq:AB oscillation }
|\Psi_t(B)|^2 \propto \, |t_{hp}(\theta)|^2 
\Bigg\{
1 + \cos\Big[\phi_{B}^L(\theta_i^{\mathrm{max}},B)+\phi_{B}^{R}(-\theta_\mathrm{max},B) 
-\phi_{B}^L(-\theta_i^{\mathrm{max}},B)-\phi_{B}^{R}(\theta_\mathrm{max},B)\Big] \Bigg\}\, ,
\end{align}
\end{widetext}
where $\theta_i^{\mathrm{max}}$ is obtained by solving Eq.~\eqref{eq:final angle} with $\theta_f=\theta_\mathrm{max}$.

Finally, as the corresponding trajectories in the $p$- and $n$- doped regions at $\pm\theta_{\mathrm{max}}$ enclose an area $A(\theta_{\rm max})$, the total Aharanov-Bohm phase in Eq.~\eqref{eq:AB oscillation } can be approximated as the magnetic flux threading through the area enclosed by the aforementioned trajectories. Moreover, as the deformation of the enclosed area remains small in the chosen range of the magnetic field, we assume the area to be independent of the field, i.e.
\begin{equation}\label{eq: density modulation}
    |\Psi_t(B)|^2 \propto |t_{hp}(\theta)|^2 \left[ 1+\cos\Big(\frac{e}{\hbar}B \, A(\theta_\mathrm{max})\Big) \right] \, ,
\end{equation}
where $A(\theta_\mathrm{max})$ is obtained using Eq.~\eqref{eq: enclosed area}. Hence, we expect the period of oscillations in conductance to be determined by the maximal transmitted angle. This agrees well with the numerical results obtained using Kwant~\cite{groth_kwant_2014}, as discussed in the main text.

\section{AB interferometer with a spatially-confined flux tube}
In Fig.~\ref{fig: setup and result1}(b) of the main text, we observe a weak dependence of the period of oscillations on the strength of the magnetic field. This is evident both from $G_B$, where the distance between maxima of oscillations become closer at higher $B$, which manifests as a wide peak in the Fourier transform.
The origin of this dependence is the small change of $A(\theta_\mathrm{max})$ with increasing $B$ due to stronger bending of trajectories at higher magnetic fields. 

To confirm the above claim, we alter our setup and conduct a numerical simulation in which the magnetic field does not pierce the whole sample, but is instead confined to a constant circular area of a radius $r_\mathrm{hole}=50\,\mathrm{nm}$ located in the $n$-doped region. Furthermore, a high onsite potential of $\mu=10^{12}\,\mathrm{meV}$ is applied inside this region, creating an effective cavity in the junction where no state can enter, see Fig.~\ref{fig: perfect AB interferometer}(a). 

Experimentally, this situation corresponds to a solenoid with a running current being placed in the $n$-doped region of our $pn$ junction.
Note that in this case, the trajectories around the maximal transmitted angle $\theta_\mathrm{max}$ will not bend at any magnetic field, and only the AB phase contributes to a total phase in Eq.~\eqref{eq:total phase}. 

For this new setup, we numerically calculate the conductance between the leads S and D for the magnetic field varying between $8$ and $12\,\mathrm{T}$. The result is shown in Fig.~\ref{fig: perfect AB interferometer}(b). From a Fourier transform shown in the lower panel, we observe that the main period of the oscillation is $\Delta B=1/1.81\approx0.55\,\mathrm{T}$, which agrees well with the period predicted by Eq.~\eqref{eq: density modulation} $h/(e\pi r_\mathrm{hole}^2)\approx 0.53\,\mathrm{T}$. We attribute the discrepancy to the finite resolution of the numerical simulation in the magnetic field. Moreover, as the simulated source lead is not a point-like object but has a finite width, the trajectories emitted from different positions in the $y$-direction can interfere with each other. We hence find higher order frequencies displayed in Fig.~\ref{fig: perfect AB interferometer}(b).

\section{Particle-like spinor injecting case}
\begin{figure}[!t]
		\centering
		\includegraphics{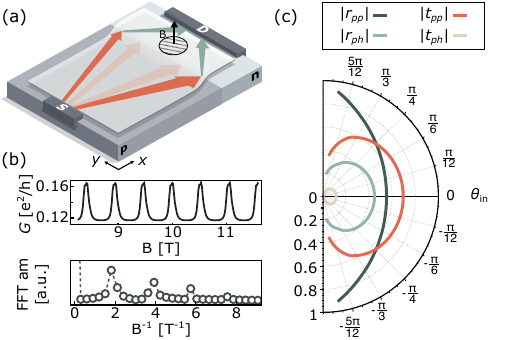}
    ~		\caption{The impact of a flux solenoid and particle-like scattering. (a) A hole is punctuated at $(x,y)=(860,0)$ with a radius $r_\mathrm{hole}=50\,\mathrm{nm}$, marked by the dashed line. A magnetic field is applied perpendicularly within the hole. (b) Upper panel: the conductance obtained numerically between S and D as a function of the magnetic field. Lower panel: The corresponding distribution of AB-oscillation  frequencies. (c) Scattering amplitudes as a function of the incident angle for an incident particle-like spinor. }
    \label{fig: perfect AB interferometer}
\end{figure}

In the main text, we restrict our discussion to the case where only a hole-like spinor is injected from the source lead, as the relevant hole-to-particle transmission supports strong negative refraction needed for realizing effective spatial confinement for electronic trajectories, which is crucial for the AB effect. Yet, in  experiments as well as the numerical simulations shown in the main text, electrons are injected from a metallic lead and are scattered into a superposition of hole-like and particle-like spinors. Therefore, for completeness, we demonstrate the behavior of the injected particle-like spinors in this section.

Similar to the previous case of the hole-like injected spinor, we calculate the respective scattering amplitudes in Eq.~\eqref{eq:scattering amplitudes} by solving a set of equations similar to Eq.~\eqref{eq: psiL R}, but for the incident particle-like spinor. With this, we obtain $r_{pp}$, $r_{ph}$, $t_{pp}$ and $t_{ph}$ as a function of the incident angle $\theta_{in}$, see Fig.~\ref{fig: perfect AB interferometer}(c). 
We find that (i) the particle-to-hole transmission, which also exhibits negative refraction~\cite{zhao_electron_2023}, has little contribution to the total conductance due to its small amplitude at all incident angles, and (ii) unlike the amplitudes presented in Fig.~\ref{fig: scattering and phase}(c), the scattering amplitudes in Fig.~\ref{fig: perfect AB interferometer}(c) exhibit a homogeneous angular distribution without a large maxima at finite angles, i.e., unlike the case for $|t_{hp}|$. 
Therefore, we expect the contribution from the injected particle-like spinor to the conductance to yield a smooth background that does not strongly depend on the magnetic field. 

\section{Angle of the maximal transmission  \label{sec: maximal transmission}}


In the following, we further investigate the relation between the angular distribution of $t_{hp}$ and the doping level of the junction as well as the band structure of the system. 
To compare the behavior of $t_{hp}$ with the band structure given in the momentum space, we label each trajectory incident to the $pn$ junction with the momentum $k_y$. The connection between $k_y$ and the incident angle is given by the relation $\theta_{\mathrm{in}}=-\sin^{-1}(k_y/k^L)$, where $k^L$ is the Fermi momentum of the hole-like branch in the $p$-doped region on the left-hand side from the $pn$ interface.

\subsection{Symmetric junction}
\begin{figure}[t!]
		\centering
		\includegraphics{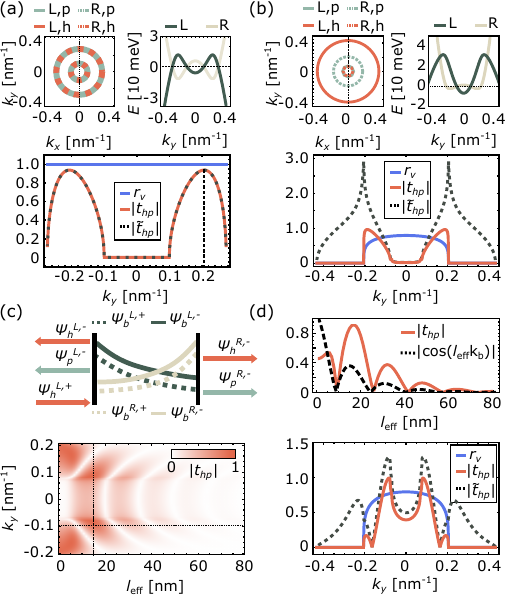}
		\caption{Junction asymmetry and graded interface as a $p-i-n$ junction. 
        (a) Scattering in the symmetric junction. Upper left: Fermi surfaces of the $p$-doped region at the left-hand side (L) and $n$-doped region at the right-hand side (R). The particle- and hole-like branches are coloured using green and red, respectively. Upper right: Dispersion in the $p$- and $n$ doped region. Lower panel: The unnormalized and normalized transmission amplitudes, $\tilde{t}_{hp}$ and $t_{hp}$, respectively, and the ratio of the velocities $r_v=\sqrt{v^R_p/v^L_h}$. The calculation is conducted using the Hamiltonian in Eq.~\eqref{eq:H continuous} with $\mu_R=-\mu_L=28\,\mathrm{meV}$ and $\mathcal{M}_2=0\,\mathrm{meV}\,\mathrm{nm}^2$. The dashed line marks $k_\mathrm{max}=\sqrt{M_0/M_2}\approx0.21\,\mathrm{nm}^{-1}$.
        (b) Scattering in an asymmetric junction. The parameters are the same as in the main text. 
        (c) Scattering in the two-interface model. The barrier region is realized by replacing $\mu_{L/R}$ in Eq.~\eqref{eq:H continuous} with $\mu_b=15\,\mathrm{meV}$ -- causing the Fermi level ro reside inside the gap. Upper panel: Illustration of the scattering processes where evanescent waves inside the barrier are labelled by $\psi_b^{L/R,\pm}$. Lower panel: The transmission amplitude $|t_{hp}|$ as a function of $k_y$ and the length of the barrier $l_\mathrm{eff}$. 
        (d) Upper panel: The amplitude of the evanescent wave inside the barrier $|\cos(k_b l_\mathrm{eff})|$, compared with the transmission amplitude $t_{hp}$ at $k_y=-0.1$nm$^{-1}$. The latter is marked by the horizontal dashed line in the lower panel of (c). Lower panel: Velocity ratio $r_v$ and (un)normalized transmission amplitude $(\tilde{t}_{hp}) t_{hp}$ as a function of $k_y$ at $l_\mathrm{eff}=15\,\mathrm{nm}$. The chosen values of $k_y$ and $l_\mathrm{eff}$ are marked by the vertical dashed line in the lower panel of (c).}
    \label{fig: maximal transmitted angle and double barrier calculation}
\end{figure}

We start with the Hamiltonian that preserves particle-hole symmetry, i.e. we set $\mathcal{M}_1=0$ in Eq.~\eqref{eq:H continuous} in the main text. Furthermore, we dope the left and right-hand side symmetrically ($\mu_L=-\mu_R$) such that the bands in the $p$- and $n$-doped regions are symmetric with respect to the Fermi level, see the upper panel of Fig.~\ref{fig: maximal transmitted angle and double barrier calculation}(a). In this case, $t_{hp}=\tilde{t}_{hp}$ since the velocity ratio $r_v=\sqrt{v^R_p/v^L_h}=1$ for all $k_y$. Scattering in such a junction was discussed in a previous work~\cite{zhao_electron_2023}. As shown in the lower panel of Fig.~\ref{fig: maximal transmitted angle and double barrier calculation}(a), the transmission amplitude $|t_{hp}|$ maximizes at $k_y=\pm k_\mathrm{max}=\pm\sqrt{-M_0/M_2}$, cf.~Ref.~\onlinecite{zhao_electron_2023} for more details. We notice that this occurs at the Fermi momentum where the particle- and hole-like dispersions cross in the case of no hybridization, i.e., at $\mathcal{A}=0$. Moreover, in this case, the transmission of different $k_y$ modes is distributed in a wide region of $k_y$-s located between the inner and the outer branch of the Fermi surface, which corresponds to a wide range of incident angles. In this case, we do not expect AB oscillations to appear due to the wide angular distribution of $|t_{hp}|$.

%
%

\subsection{Asymmetric junction}
Once the junction is doped to different chemical potentials on each side, or the particle-hole symmetry is broken by a finite value of $\mathcal{M}_1$, the ratio of the velocity $r_v$ varies as $k_y$ changes.
Moreover, the maximum of transmission is shifted away from the $k_{\rm max}$ defined above. 
Nevertheless, a general form for the scattering amplitude reads
\begin{equation}\label{eq: thp ansatz}
\footnotesize
\begin{split}
    \tilde{t}_{hp}&=\frac{c_1(k^L_pk^L_h+k^L_hk^R_h)}
    { \left[c_2(k^L_pk^L_h+k^R_pk^R_h)+c_3(k^L_pk^R_p+k^L_hk^R_h)-c_4(k^L_hk^R_p+k^L_pk^R_h)\right]} \, ,
\end{split}
\end{equation}
where $c_{1}$, $c_{2}$, $c_{3}$ and $c_{4}$ do not depend on $k_x$ or $k_y$ momenta and are fixed by the doping level. The change of $\tilde{t}_{hp}$ as a function of $k_y$ is hence determined by the behavior of the $k_x$ momentum of spinors, which we denote as $k^{L/R}_{h/p}$, see also Sec.~\ref{app:scattering matrix approach}.
We take real $k^{L/R}_{h/p}=|k^{L/R,\pm}_{h/p}|$ for propagating spinors [i.e. $\mathrm{Im}(k^{L/R,\pm}_{h/p})=0$], and imaginary $k^{L/R}_{h/p}=i|\mathrm{Im}(k^{L/R,\pm}_{h/p})|$ for spinors bound at the $pn$ interface. 

For the asymmetric junction discussed in the main text, also shown in Fig.~\ref{fig: maximal transmitted angle and double barrier calculation}(b), $\tilde{t}_{hp}$ reaches its maximum around the point where $k_x$-momentum of the particle-like spinor in the $n$-doped region vanishes, i.e. $k^{R}_p=0$. This follows directly from the expression in Eq.~\eqref{eq: thp ansatz}. 
Namely, we first notice that when both types of spinors are present on both sides of the junction, two momenta located at the inner branch, i.e. $k^L_p$ and $k^R_h$ have a stronger dependency on $k_y$ than the momenta located at the outer branch, i.e. $k^L_h$ and $k^R_p$. Therefore, the behavior of $\tilde{t}_{hp}$ at small $k_y$ is governed mostly by $k^L_p$ and $k^R_h$ in the numerator. 
After $k_y$ passes the termination points of the inner branches, $k^L_p$ and $k^R_h$ become monotonously increasing imaginary functions of $k_y$ (this follows from solving $k_x$ through analytic continuation in the complex plane).
This results in an increasing numerator in Eq.~\eqref{eq: thp ansatz}, while the increase in the denominator is suppressed by the reduction of $k^R_p$ lying on the outer Fermi surface in the $n$-doped region so the $|\tilde{t}_{hp}|$ monotonously grows. 
Furthermore, the reduction of $k^{R}_p$ results in a reduction of the velocity ratio $r_v$ until it reaches zero at $k_y$ that gives $k^{R}_p(k_y)=0$, see the blue line in the lower panel of Fig.~\ref{fig: maximal transmitted angle and double barrier calculation}(b). The normalized transmission amplitude $|t_{hp}|$, therefore, maximizes close to the vanishing point of $k^{R}_p$, see the lower panel of Fig.~\ref{fig: maximal transmitted angle and double barrier calculation}(b). 

Moreover, note that $|t_{hp}|$ has a relatively large value only in a narrow range of $k_y$ located between the points where the inner Fermi surface of the $p$-doped region and the outer Fermi surface of the $n$-doped region vanish.
This range can be tuned by controlling the relative size of the Fermi surfaces, which is done by tuning $\mu_{L/R}$. Therefore, an effective spatial confinement in our system is controllable by changing the doping level of the system.

\section{Scattering at the graded interface}
\subsection{Two-interface approach}
The scattering amplitude through the graded interface can be calculated using a transfer matrix method, cf. Ref.~\cite{cheianov_selective_2006}. Here, we use a two-interface model to describe the barrier that emerges once the chemical potential is inside the band gap. In other words, the graded interface is replaced by a gapped region having an effective length $l_{\mathrm{eff}}$ in $x$-direction, see the upper panel in Fig.~\ref{fig: maximal transmitted angle and double barrier calculation}(c). The gapped-region is realized by replacing $\mu_{L/R}$ with $\mu_b=15\,\mathrm{meV}$ in Eq.~\eqref{eq:H continuous} -- a value that sets the Fermi level to the middle of the energy gap. Solving Hamiltonian in this gapped region yields four spinors $\psi_b^{L/R,\pm}$, localized at the left (L) and right (R) interface. The corresponding four $k_x$ momenta acquire the same absolute value i.e. $k^{L/R,\pm}_b=\pm|\mathrm{Re}(k_b)|\pm|\mathrm{Im}(k_b)|$, with $k_b$ obtained from the equation $E(k_b,k_y)=0$. Note that the superscript $\pm$ in $k_b^{L,R,\pm}$ does not carry the same physical interpretation as before where it denoted the propagating direction towards the interface, since the group velocity is not defined in the band gap. The total $S$-matrix incorporating this gapped region is determined by the continuity conditions on the two interfaces
\begin{equation}\label{eq: continuity condition2}
    \begin{split}
        \Psi^{L}(\mathbf{x},k)\big|_{x=-l_\mathrm{eff}/2}&= \Psi^{m}(\mathbf{x},k)\big|_{x=-l_\mathrm{eff}/2},\\
        \partial_x\Psi^{L}(\mathbf{x},k)\big|_{x=-l_\mathrm{eff}/2}&= \partial_x\Psi^{m}(\mathbf{x},k)\big|_{x=-l_\mathrm{eff}/2},\\
        \Psi^{m}(\mathbf{x},k)\big|_{x=l_\mathrm{eff}/2}&= \Psi^{R}(\mathbf{x},k)\big|_{x=l_\mathrm{eff}/2},\\
        \partial_x\Psi^{m}(\mathbf{x},k)\big|_{x=l_\mathrm{eff}/2}&= \partial_x\Psi^{R}(\mathbf{x},k)\big|_{x=l_\mathrm{eff}/2},
    \end{split}
\end{equation}
where $\Psi^{L/R}$ is the wavefuncion in the $p$- and $n$-doped regions defined in Eq.~\eqref{eq: psiL R}. The wavefuncion in the gapped region is given by
\begin{align}
    \Psi^{m}(\mathbf{x},k)=&a \psi^{L,+}_{b}e^{-i\mathbf{k}_{h}^{L,+}\mathbf{x}}+b \psi^{L,-}_{b}e^{-i\mathbf{k}_{h}^{L,-}\mathbf{x}} \nonumber\\
    &+c \psi^{R,+}_{b}e^{-i\mathbf{k}_{h}^{R,+}\mathbf{x}}+d \psi^{R,-}_{b}e^{-i\mathbf{k}_{h}^{R,-}\mathbf{x}} \, .
\end{align}
We insert $\Psi^{m}$ into Eq.~\eqref{eq: continuity condition2} and solve it to obtain the scattering amplitudes $a, b, c$ and $d$ for the in-gap states as well as the amplitudes for states in the $n$-doped region $\tilde{t}_{hp}$ and $\tilde{t}_{hh}$. In the lower panel of Fig.~\ref{fig: maximal transmitted angle and double barrier calculation}(c), we show the obtained hole-to-particle transmission amplitude $|t_{hp}|$ as a function of the length of the gapped region $l_\mathrm{eff}$ and $k_y$. Periodic resonances of $|t_{hp}|$ appear as a function of $l_\mathrm{eff}$ and we attribute them to the in-gap resonances caused by the  evanescent waves. Furthermore, in the upper panel of Fig.~\ref{fig: maximal transmitted angle and double barrier calculation}(d), we show $|t_{hp}|$ at constant $k_y=-0.1\,\mathrm{nm}^{-1}$, and compare it with the amplitude $\left| \cos(k^{L,+}_b l_\mathrm{eff}) \right|$ of the evanescent wave inside the gap. We find that the transmission amplitude $|t_{hp}|$ follows a qualitatively identical behavior as the amplitude of the evanescent wave; with the maxima and minima being at points determined by the standing modes of the evanescent wave. The overall decay of $|t_{hp}|$ with increasing $l_\mathrm{eff}$ is as well captured by the decreasing amplitudes of the evanescent waves.

\subsection{A change in the maximal transmission with increasing $l_\mathrm{eff}$}
Let us now turn to the $k_y$ dependence of $|t_{hp}|$ when $l_\mathrm{eff}$ is changed.  As shown in the lower panel of Fig.~\ref{fig: maximal transmitted angle and double barrier calculation}(c), for small $l_\mathrm{eff}$, $|t_{hp}|$ maximizes at $k_y\approx\pm0.2$nm$^{-1}$, see also Sec.~\ref{sec: maximal transmission}. However, when $l_\mathrm{eff}$ approaches $l_\mathrm{eff}\approx 10\,\mathrm{nm}$, another local maximum appears at $k_y \approx \pm 0.07\,\mathrm{nm}^{-1}$, which then grows as $l_\mathrm{eff}$ is increased and eventually becomes a new global maximum of $|t_{hp}|$ at $l_\mathrm{eff}\approx 15\,\mathrm{nm}$, see the lower panel of Fig.~\ref{fig: maximal transmitted angle and double barrier calculation}(d). 
The former occurs at $l_\mathrm{eff}$ that is smaller than the decaying length of the evanescent waves and, therefore, allows a higher-order scattering process to happen inside the gapped region. 
This leads to a strong hybridization of the evanescent states emanating from the two interfaces. The barrier in this case is relatively “transparent”, and $|t_{hp}|$ has a maximum at the same spot as in the case without the barrier. As $l_\mathrm{eff}$ increases beyond the decay length of the evanescent waves, the contribution to the tunneling is dominated by the scattering between $p$($n$)-doped and the gapped region. We argue that for this case, the transmission is facilitated mostly by the vanishing of $k^L_p$, and hence the high peaks occur at $k^{L}_p=0$.

\end{document}